\documentclass[11pt]{article}

\usepackage{graphicx}        

\usepackage{axodraw}

\newcommand{\Li}[2]{{\mbox{Li}}_{#1}\left(#2\right)}

\newcommand{\veps}{\varepsilon}

\newcommand{\ha}{\frac{1}{2}\:}
\newcommand{\wz}{\sqrt{2}}

\newcommand{\nn}{\nonumber}
\newcommand{\crn}{\nonumber \\}
\newcommand{\be}{\begin{equation}}
\newcommand{\ee}{\end{equation}}
\newcommand{\bea}{\begin{eqnarray}}
\newcommand{\eea}{\end{eqnarray}}
\newcommand{\ba}{\begin{eqnarray*}}
\newcommand{\ea}{\end{eqnarray*}}
\newcommand{\lbl}[1]{\label{eq:#1}}

\newcommand{\VVdA}{\langle VV\partial A \rangle}

\newcommand{\epo}{\;\:.}
\newcommand{\amu}{a_\mu}

\begin{document}

\begin{titlepage}

\begin{flushright}
\today  \\
HU-EP-05-66~~~\\
DESY-05-211~~~\\
SFB/CPP-05-69\\
\end{flushright}

\vspace*{0.2cm}
\begin{center}
{\Large {\bf Explicit results for the anomalous three point function
\\[0.5cm]
and  non-renormalization theorems 
}}\\[2 cm]
{\bf F.~Jegerlehner}$^a$ {\bf and  O.V.~Tarasov}$^b$\\[1cm]

$^a$ {\it Humboldt Universit\"at zu Berlin, Institut f\"ur Physik\\
       Newtonstrasse 15, D-12489 Berlin, Germany\\
       E-mail: {\tt fjeger@physik.hu-berlin.de}} \\

$^b$  {\it Deutsches Elektronen Synchrotron DESY\\
       Platanenallee 6, D-15738 Zeuthen, Germany\\
     E-mail: {\tt Oleg.Tarasov@desy.de}}\\

\end{center}

\vspace*{1.0cm}

\begin{abstract}
Two-loop corrections for the $\langle VVA \rangle$ correlator of the
singlet axial and vector currents in QCD are calculated in the chiral
limit for arbitrary momenta.  Explicit calculations confirm the
non-renormalization theorems derived recently by Vainshtein [Phys.\
Lett.\ B {\bf 569} (2003) 187] and Knecht et al. [JHEP {\bf 0403}
(2004) 035]. We find that as in the one-loop case also at two loops
the $\langle VVA \rangle$ correlator has only three independent
form-factors instead of four. From the explicit results we observe
that the two-loop correction to the correlator is equal to the
one-loop result times the constant factor $C_2(R) \alpha_s/\pi$ in the
$\overline{\rm MS}$ scheme.  This holds for the full correlator, for
the anomalous longitudinal as well as for the non-anomalous
transversal amplitudes. The finite overall $\alpha_s$ dependent
constant has to be normalized away by renormalizing the axial current
according to Witten's algebraic/geometrical constraint on the
anomalous Ward identity [$\langle VV\partial A \rangle$ correlator].
Our observations, together with known facts, suggest that in
perturbation theory the  $\langle VVA \rangle$ correlator is
proportional to the one-loop term to all orders and that the
non-renormalization theorem of the Adler-Bell-Jackiw anomaly carries
over to the full correlator.

\end{abstract}

\end{titlepage}

\section{Introduction}
The Adler-Bell-Jackiw ~\cite{Adler69,BellJackiw69} triangle anomaly in
the divergence of the axial vector current is well known to play a
crucial role at several places in elementary particle physics. Its
nature is controlled by the Adler-Bardeen non-renormalization
theorem~\cite{AdlerBardeen69}, by the Wess-Zumino integrability
condition and the Wess-Zumino effective action~\cite{WessZumino71}, by
Witten's algebraic/geometrical interpretation which requires the axial
current to be normalized to an integer~\cite{Witten83} and by the
t'~Hooft quark-hadron duality matching
conditions~\cite{tHooft79}. Phenomenologically, it plays a key role in
the prediction of $\pi^0 \to \gamma \gamma$, and in the solution of
the $U(1)$ problem.  Last but not least, renormalizability of the
electroweak Standard Model requires the anomaly cancellation which
dictates the lepton-quark family structure.

More recently Vainshtein~\cite{Vainshtein03} found an important new
relation $w_T(q^2)=\ha w_L(q^2)$ (see below) matching to all orders in
perturbation theory, in some kinematical limit, the transversal part
to the anomalous longitudinal amplitude which is subject to the
Adler-Bardeen non-renormalization theorem. Later Knecht et
al.~\cite{KPPdR04} were able to generalize this kind of
non-renormalization theorems.  These recent investigations came up in
connection with problems in calculating the leading hadronic
effects in the electroweak two--loop contributions to the muon 
anomalous magnetic moment $a_\mu$~\cite{PPdR95,CKM95,KPPdR02,CMV03}.

The first electroweak two--loop calculations~\cite{KKSS92} for $a_\mu$
revealed that triangle fermion--loops may give rise to unexpectedly
large radiative corrections. The diagrams which yield the leading
corrections are those including a VVA triangular fermion--loop
($VVA\neq0$ while $VVV=0$ ) associated with a $Z$ boson exchange \\[-6mm]
\begin{center}
\includegraphics[height=2.5cm]{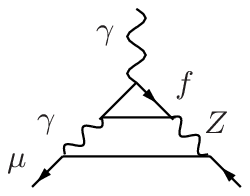}
\end{center}
and a fermion of flavor $f$ gives a contribution, up to UV singular terms which
will cancel,
\ba
a^{(4)\:\mathrm{EW}}_\mu([f])
\simeq \frac{\wz G_\mu m_\mu^2}{16 \pi^2}
\frac{\alpha}{\pi}\:2T_fN_{cf}Q_f^2\:\left[3 \ln \frac{M_Z^2}{m_{f'}^2} + C_f
\right]
\ea
where $\alpha$ is the fine structure constant, $G_\mu$ the Fermi
constant, $T_{3f}$ the 3rd component of the weak isospin, $Q_f$ the
charge and $N_{cf}$ the color factor, 1 for leptons, 3 for quarks. The
mass $m_{f'}$ is $m_\mu$ if $m_f < m_\mu$ and $m_f$ if $m_f >
m_\mu$. $C_f$ denotes constant terms.  Since, as granted in the
Standard Model of elementary particles, anomaly cancellation by
lepton--quark duality $\sum_fN_{cf}Q_f^2T_{3f}=0$ is at work, only the
sums over complete lepton--quark families yield meaningful results
relevant to physics.  In any case the quark contributions have to be taken
into account. In fact treating the quarks like free fermions (quark
parton model QPM) the first family yields $$
\amu^{(4)\:\mathrm{EW}}([e,u,d])_\mathrm{QPM}\simeq
-\frac{\wz G_\mu\:m_\mu^2}{16 \pi^2}\:\frac{\alpha}{\pi}\:\left[
\ln \frac{m_u^8}{m_\mu^6m_d^2}+\frac{17}{2}\right]
$$ which demonstrates that the leading large logs $\sim \ln M_Z$ have
dropped! However, the quark masses which appear here are illdefined
constituent quark masses, which can hardly account reliably for the strong
interaction effects.

Since we are interested in a static low energy quantity $a_\mu=\ha
(g-2)_\mu=F_\mathrm{M}(0)$, given by the Pauli form factor at zero
momentum transfer, the perturbative QCD (pQCD) calculation of the
light quark contributions seems more than questionable. However,
there is a large scale in the game, namely the $Z$ boson mass $M_Z\sim
91.19$ GeV, which makes an estimate by the quark parton model as a
first step not completely unreasonable. Indeed, in the relevant
kinematical region, the leading strong interaction effects may be
parametrized by two VVA amplitudes, a longitudinal $w_L(Q^2)$ and a
transversal $w_T(Q^2)$ one, which contribute as~\cite{KPPdR02,CMV03}
\bea
\Delta a_\mu^{(4)\:\mathrm{EW}}([f])_{\mathrm{VVA}}&\simeq&
\frac{\wz G_\mu\:m_\mu^2}{16 \pi^2}\:\frac{\alpha}{\pi}\:
\int_{m_\mu^2}^{\Lambda^2}\:d Q^2\:\left(
w_L(Q^2)+ \frac{M_Z^2}{M_Z^2+Q^2}\:w_T(Q^2)\right)\;,
\eea 
where $\Lambda$ is a cutoff to be taken to $\infty$ at the end.
For a perturbative fermion loop to leading order~\cite{Rosenberg63}
\bea
w_L^{1-\mathrm{loop}}(Q^2)=2
w_T^{1-\mathrm{loop}}(Q^2)&=&\sum_f\:4T_fN_{cf}Q_f^2\:
\int_0^1 \frac{d x\:x\:(1-x)}{x\:(1-x)\:Q^2+m_f^2} \nonumber \\
&\stackrel{m^2_f \ll Q^2}{=}&
\sum_f\:4T_fN_{cf}Q_f^2\:\left[\frac{1}{Q^2}-\frac{2m_f^2}{Q^4}\:\ln \frac{Q^2}{m_f^2}
+O(\frac{1}{Q^6}) \right] \epo
\eea
Vainshtein~\cite{Vainshtein03} has shown that in the chiral limit the relation 
\bea
\left. w_T(Q^2)_\mathrm{pQCD}\right|_{m=0}=\ha 
\left. w_L(Q^2)\right|_{m=0}
\label{nonrentrans}
\eea
is valid actually to all orders of perturbative QCD. Thus the
non-renormalization theorem valid beyond pQCD for the anomalous
amplitude $w_L$ (considering the quarks $q=u,d,s,c,b,t$ only):
\bea 
\left. w_L(Q^2)\right|_{m=0}=
\left. w^{1-\mathrm{loop}}_L(Q^2)\right|_{m=0}=  \sum_q (2T_qQ_q)\:  \frac{2 N_c}{Q^2}
\eea
carries over to the perturbative part of the transversal amplitude.
Thus in the chiral limit the perturbative QPM result for $w_T$ is 
exact. This may be somewhat puzzling, since in low energy effective QCD,
which encodes the non-perturbative strong interaction effects, 
this kind of term seems to be absent.

We observe that the contributions from $w_L$ for individual fermions
is logarithmically divergent, but it completely drops for a complete
family due to the vanishing anomaly cancellation coefficient. The
contribution from $w_T$ is convergent for individual fermions due to
the damping by the $Z$ propagator. In fact it is the leading $1/Q^2$
term of the $w_T$ amplitude which produces the $\ln \frac{M_Z}{m}$
terms. However, the coefficient is the same as for the anomalous term
and thus for each complete family also the $\ln M_Z$ terms must drop
out.

Low energy QCD is characterized in the chiral limit of massless light
quarks $u,d,s$, by spontaneous chiral symmetry breaking (S$\chi$SB) of
the chiral group $SU(3)_V \otimes SU(3)_A$, which in particular
implies the existence of the pseudoscalar octet of pions and kaons as
Goldstone bosons. The light quark condensates are essential features in
this situation and lead to non-perturbative effects completely absent
in a perturbative approach. Thus such low energy QCD effects are
intrinsically non--perturbative and controlled by chiral perturbation
theory ($\chi$PT), the systematic QCD low energy expansion, which
accounts for the S$\chi$SB and the chiral symmetry breaking by quark
masses in a systematic manner. The low energy effective theory
evaluation of the hadronic contributions related to the light quarks
$u,d,s$ was worked out in~\cite{PPdR95,KPPdR02} and later
in~\cite{CMV03}. It was shown that in the operator product
expansion (OPE) the leading non-perturbative (NP) term in the chiral
limit is due to the $u,d,s$ quark condensate $\langle
\bar{\psi}\psi \rangle \neq 0$
\bea
w_T(Q^2)_\mathrm{NP} \simeq \frac{16}{9}\pi^2\:\frac{1}{M_\rho^2}\:
\frac{\alpha_s}{\pi}\frac{\langle \bar{\psi}\psi \rangle^2}
{Q^6}~~\mathrm{at}~~Q^2~~\mathrm{large}
\eea
which breaks the degeneracy $w_T(Q^2)=\ha w_L(Q^2)$ found in 
perturbation theory\footnote{The OPE only provides information on
$w_T$ for $Q^2$ large. At low $Q^2$ we only know that 
$w_T(0)=128 \pi^2\:C^W_{22}$ where $C^W_{22}$ is one of the unknown 
$\chi$PT
constants in the $O(p^6)$ parity odd part of the chiral 
Lagrangian~\cite{BGT02}. The low energy effective theory does not
yield a $1/Q^2$--pole as required by the non-renormalization theorem. 
In the OPE $1/Q^4$ terms are due to explicite chiral
symmetry breaking which yields $\Delta w_T(Q^2)_\mathrm{NP}\simeq -\frac{4}{9}\:\frac{1}{M_\rho^2}\:\frac{(4m_u-m_d-m_s)
\langle \bar{\psi}\psi \rangle}{Q^4}$.}. $M_\rho$ is the $\rho$ mass.

Relevant effective couplings are the neutral current part   
$$ {\cal L}^{(2)}=- \frac{e}{2\sin
\Theta_W \cos \Theta_W} f_\pi \partial_\mu
\left(\pi^0+\frac{1}{\sqrt{3}}\:\eta_8-\frac{1}{\sqrt{6}}\:\eta_0
\right)\:Z^\mu \;,
$$
and the Wess-Zumino Lagrangian
$$
{\cal L}_\mathrm{WZ}=\frac{\alpha}{\pi} \frac{N_c}{24
f_\pi}\:\left(\pi^0+\frac{1}{\sqrt{3}}\:\eta_8+2
\sqrt{\frac{2}{3}}\:\eta_0  \right)\:\veps_{\mu \nu \rho \sigma} F^{\mu
\nu }F^{\rho \sigma}\;,
$$  
which reproduces the ABJ anomaly via the PCAC relation. $e$ is the
positron charge, $\sin^2 \Theta_W$ is the weak mixing parameter,
$f_\pi$ the pion decay constant and $\pi^0$ is the neural pion field. 
The pseudoscalars $\eta_8,\eta_0$ are mixing to $\eta,\eta'$.
The $[u,d,s]$ contribution evaluated this way is given by the diagram
\begin{center}
\includegraphics[height=2cm]{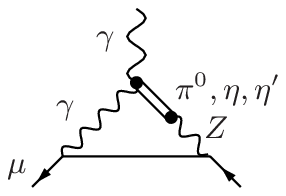}
\end{center}
which together with its crossed version in the unitary gauge and in
the chiral limit and completed for the first two fermion families 
yields~\cite{PPdR95,KPPdR02}
\ba 
\amu^{(4)\:\mathrm{EW}}([e,u,d;\mu,c,s])_{\chi\mathrm{PT}}&=&
\frac{\wz G_\mu\:m_\mu^2}{16 \pi^2}\:\frac{\alpha}{\pi}\:\left[
-\frac{14}{3} \ln \frac{M_Z^2}{m_\mu^2}+4\ln \frac{M_Z^2}{m_c^2}-\frac{35}{3}
+\frac{8}{9}\pi^2\right] \;.
\ea
The residual $M_Z$--dependence in effective theory was
controversial\footnote{Numerical estimates presented in~\cite{KPPdR02}
and~\cite{CMV03} agree well within errors and the discrepancy in
question is of conceptual nature and not relevant for the
interpretation of the present $g-2$ experiment.}, and is contradicting
Vainshtein's non-renormalization theorem~\cite{Vainshtein03,CMV03}. In
fact, there is not really a contradiction. The point is that the low
energy effective theory does not apply up to the $M_Z$ mass scale as
assumed in obtaining the above result. One rather has to apply a
cut--off $\Lambda_\mathrm{matching}$ of the order of the proton mass
$m_p$, say, and above that scale of course the QPM gives the correct
answer. The leading large log terms proportional to $\ln M_Z$ then again cancel and the
$M_Z$--dependence gets replaced by a
$\Lambda_\mathrm{matching}$--dependence (with coefficient -14/3
changed to -2/3 and -35/3 to -47/3) in the above result.

As the extensions of the Adler-Bardeen non-renormalization theorem
for the anomalous Ward identity $\VVdA$ turn out to play an
important role in new phenomenological applications, we will study in
the following such possible generalizations by an explicit
calculation of the leading QCD corrections to the $\gamma \gamma Z$
triangle.

\section{Definitions}
Let us consider the VVA three point function
\be
\label{Greencalw}{\cal W}_{\mu\nu\rho}(q_1,q_2) = i\int d^4x_1 d^4x_2
\,e^{i(q_1\cdot x_1 + q_2\cdot x_2)}
\,\times\,\langle\,0\,\vert\,\mbox{T}\{V_\mu(x_1)V_\nu(x_2)A_\rho(0)\}
\,\vert\,0\,\rangle
\ee 
of the flavor and color diagonal fermion currents 
\be V_{\mu} \,=\,{\overline\psi}\gamma_\mu\,\psi \quad ,
\quad A_{\mu} \,=\,
{\overline\psi}\gamma_\mu\gamma_5\,\psi 
\ee
where $\psi$ is a quark field. The vector currents are strictly conserved $\partial_\mu
V^\mu=0$, while the axial vector current satisfies a PCAC relation
plus the anomaly $\partial_\mu A^\mu=2im_0 \bar{\psi}\gamma_5 \psi
+\frac{\alpha_0}{4\pi}\veps_{\mu \nu \rho \sigma}F^{\mu \nu}(x)F^{\rho
\sigma}(x)$.  
We will be mainly interested in the properties of strongly interacting quark 
flavor currents in perturbative QCD. To leading order the correlator
of interest is associated with the one--loop triangle diagram 
\begin{center}
\begin{picture}(60,40)(50,160)
\ArrowLine(65,165)(105,185)
\ArrowLine(105,185)(105,145)
\ArrowLine(105,145)(65,165)
\Photon(50,165)(65,165){2}{3}
\Photon(105,185)(120,185){2}{3}
\Photon(105,145)(120,145){2}{3}
\Vertex(65,165){2}
\Vertex(105,185){2}
\Vertex(105,145){2}
\Text(53,175)[]{$A_\rho$}
\Text(114,196)[]{$V_\mu$}
\Text(114,135)[]{$V_\nu$}
\Text(140,185)[]{$\leftarrow q_1 $}
\Text(140,145)[]{$\leftarrow q_2 $}
\end{picture}
\end{center}

\vspace{9mm}

plus its crossed ($q_1,\mu \leftrightarrow q_2,\nu$) partner.
In the following we will closely follow the notation of~\cite{KPPdR04}.

\noindent  The Ward identities restrict the general covariant 
decomposition of ${{\cal W}}_{\mu\nu\rho}(q_1,q_2)$ into invariant
functions to four terms
\bea \label{calw}{{\cal W}}_{\mu\nu\rho}(q_1,q_2) &=&
 -\,\frac{1}{8\pi^2}\,\bigg\{
-w_L\left(q_1^2,q_2^2,q_3^2\right)\,(q_1+q_2)_\rho\, \veps_{\mu\nu\alpha\beta}\
q_1^\alpha q_2^\beta \nonumber\\
&& \hspace*{-3.2cm} +\,
w_T^{(+)}\left(q_1^2,q_2^2,q_3^2\right)\,t^{(+)}_{\mu\nu\rho}(q_1,q_2)
 +\,w_T^{(-)}
 \left(q_1^2,q_2^2,q_3^2\right)\,t^{(-)}_{\mu\nu\rho}(q_1,q_2)+\,
{\widetilde{w}}_T^{(-)}\left(q_1^2,q_2^2,q_3^2\right)\,{\widetilde{t}}^{(-)}_{\mu\nu\rho}(q_1,q_2)
\bigg\}\,, \eea 
with the transverse tensors given by
\bea t^{(+)}_{\mu\nu\rho}(q_1,q_2) &=&
q_{1\nu}\,\veps_{\mu\rho\alpha\beta}\ q_1^\alpha q_2^\beta \,-\,
q_{2\mu}\,\veps_{\nu\rho\alpha\beta}\ q_1^\alpha q_2^\beta \,-\, (q_{1}\cdot
q_2)\,\veps_{\mu\nu\rho\alpha}\ (q_1 - q_2)^\alpha
\nonumber\\
&& \quad\quad+\ \frac{q_1^2 + q_2^2 - q_3^2}{q_3^2}\
\veps_{\mu\nu\alpha\beta}\ q_1^\alpha q_2^\beta(q_1 + q_2)_\rho
\nonumber \ , \\
t^{(-)}_{\mu\nu\rho}(q_1,q_2) &=& \left[ (q_1 - q_2)_\rho \,-\, \frac{q_1^2 - q_2^2}{(q_1
+ q_2)^2}\,(q_1 + q_2)_\rho \right] \,\veps_{\mu\nu\alpha\beta}\ q_1^\alpha q_2^\beta
\nonumber\\
{\widetilde{t}}^{(-)}_{\mu\nu\rho}(q_1,q_2) &=& q_{1\nu}\,\veps_{\mu\rho\alpha\beta}\
q_1^\alpha q_2^\beta \,+\, q_{2\mu}\,\veps_{\nu\rho\alpha\beta}\ q_1^\alpha q_2^\beta
\,-\, (q_{1}\cdot q_2)\,\veps_{\mu\nu\rho\alpha}\ (q_1 + q_2)^\alpha \,. \lbl{tensors}
\eea 
Bose symmetry ($q_1,\mu \leftrightarrow q_2,\nu$) entails 
\bea w_T^{(+)}\left(q_2^2,q_1^2,q_3^2\right)
&=&+w_T^{(+)}\left(q_1^2,q_2^2,q_3^2\right),
\nonumber \\
w_T^{(-)}\left(q_2^2,q_1^2,q_3^2\right) &=& -
w_T^{(-)}\left(q_1^2,q_2^2,q_3^2\right),
~~~
{\widetilde{w}}_T^{(-)}\left(q_2^2,q_1^2,q_3^2\right) = -
{\widetilde{w}}_T^{(-)}\left(q_1^2,q_2^2,q_3^2\right) \,. 
\eea 
The longitudinal part is entirely fixed by the anomaly, 
\be
w_L\left(q_1^2,q_2^2,q_3^2\right)\,=\, - \frac{2N_c}{q_3^2}\,
\lbl{w_Lexpr} 
\ee 
which is exact to all orders of perturbation theory, the famous
Adler-Bardeen non-renormalization theorem.

In~\cite{KPPdR04} the following three chiral symmetry relations between amplitudes 
were derived in pQCD:
\ba &&\!\!\!\!\!\!\!\!\!\! \Biggl\{\left[w_T^{(+)}
\,+\, w_T^{(-)}\right] \left(q_1^2,q_2^2,q_3^2\right) \, -\, \left[w_T^{(+)} \,+\,
w_T^{(-)}\right] \left(q_3^2,q_2^2,q_1^2\right)\Biggr\}_{\mathrm{pQCD}}\ =\ 0
\\
&&\!\!\!\!\!\!\!\!\!\! \Biggl\{\left[{\widetilde{w}}_T^{(-)} \,+\, w_T^{(-)}\right]
\left(q_1^2,q_2^2,q_3^2\right) \, +\, \left[{\widetilde{w}}_T^{(-)} \,+\,
w_T^{(-)}\right] \left(q_3^2,q_2^2,q_1^2\right)\Biggr\}_{\mathrm{pQCD}}\  =\  0
\nonumber\\[-6mm]
\ea 
and\\[-6mm] 
\bea \label{theorems}&& \!\!\!\!\!\!\!\!\!\!\!\!\!\!\Biggl\{\left[w_T^{(+)}
\,+\, {\widetilde{w}}_T^{(-)}\right]\left(q_1^2,q_2^2,q_3^2\right) \,+\, \left[{{w}}_T^{(+)}
\,+\, {\widetilde{w}}_T^{(-)}\right] \left(q_3^2,
q_2^2,q_1^2\right)\Biggr\}_{\mathrm{\!\!pQCD}} \!\!\!\!\!\!- w_L
\left(q_3^2,q_2^2,q_1^2\right)\nonumber \\ &&  = -\Biggl\{ \frac{2\ (q_2^2 + q_1\cdot
q_2)}{q_1^2}\, w_T^{(+)} \left(q_3^2,q_2^2,q_1^2\right) \ -\, 2\,\frac{q_1\cdot
q_2}{q_1^2}\, w_T^{(-)} \left(q_3^2,q_2^2,q_1^2\right)\Biggr\}_{\mathrm{pQCD}}
\,, \eea 
involving the transverse part of the $\langle VVA\rangle$ correlator
${{\cal W}}_{\mu\nu\rho}(q_1,q_2)$, and which hold for all values of
the momentum transfers $q_1^2$, $q_2^2$ and $q_3^2$. 
In the kinematical configuration relevant for g-2 calculations,
$q_1=k \pm q,~q_2=-k$, expanding to linear order in $k$ and noting that
$\widetilde{t}^{(-)}_{\mu\nu\rho}(q_1,q_2) \sim
t^{(+)}_{\mu\nu\rho}(q_1,q_2) = q^2 \veps_{\mu \nu \rho
\sigma}k^\sigma -q_\mu \veps_{\nu \rho \alpha \beta} q^\alpha k^\beta
-q_\rho \veps_{\mu \nu \alpha \beta} q^\alpha k^\beta +O(k^2)$ and
$t^{(-)}_{\mu\nu\rho}(q_1,q_2)= O(k^2)$, these relations imply the non-renormalization
theorem (\ref{nonrentrans}) obtained in
Refs.~\cite{Vainshtein03,CMV03} upon identifying 
\bea
w_L(Q^2) \equiv w_L(-Q^2,0,-Q^2)\;,\;\;
w_T(Q^2) \equiv w_T^{(+)}(-Q^2,0,-Q^2)+\widetilde{w}_T^{(-)}(-Q^2,0,-Q^2)\;,
\eea 
with $Q^2=-q^2$.

\section{Calculations}

We perform the calculation with conventional dimensional regularization
\cite{dimreg} and use a linear covariant gauge with arbitrary gauge
parameter $\xi$ throughout the calculation.

\noindent
\vspace{40mm}
\vglue 5mm
\begin{center}
\begin{picture}(0,0)(200,55)
\Vertex(65,165){1}
\ArrowLine(65,165)(105,180)
\Vertex(105,180){1}
\Line(105,180)(105,165)
\ArrowLine(105,165)(105,150)

\Vertex(105,150){1}
\Line(82,159)(65,165)
\ArrowLine(105,150)(82,159)
\Vertex(105,170){1}
\Gluon(105,170)(80,160){2}{5}
\Vertex(80,160){1}

\Photon(55,165)(65,165){2}{3}
\Photon(105,180)(115,180){2}{3}
\Photon(105,150)(115,150){2}{3}

\Text(60,175)[]{$A_{\rho}$}
\Text(122,184)[]{$V_{\mu}$}
\Text(122,146)[]{$V_{\nu}$}

\Vertex(135,165){1}
\ArrowLine(175,180)(135,165)
\Vertex(175,180){1}
\Vertex(175,150){1}

\Line(175,180)(175,165)
\ArrowLine(175,150)(175,165)

\Photon(125,165)(135,165){2}{3}
\Photon(175,180)(185,180){2}{3}
\Photon(175,150)(185,150){2}{3}

\Line(152,159)(135,165)
\ArrowLine(152,159)(175,150)

\Vertex(175,170){1}
\Gluon(175,170)(150,160){2}{5}
\Vertex(150,160){1}

\Vertex(205,165){1}
\Line(205,165)(220,170)
\ArrowLine(220,170)(245,180)

\Vertex(245,180){1}

\ArrowLine(245,180)(245,165)
\Line(245,165)(245,150)

\Vertex(245,150){1}
\ArrowLine(245,150)(205,165)

\Vertex(220,170){1}
\Gluon(220,170)(245,162){2}{5}
\Vertex(245,162){1}

\Photon(195,165)(205,165){2}{3}
\Photon(245,180)(255,180){2}{3}
\Photon(245,150)(255,150){2}{3}

\Vertex(275,165){1}
\Line(275,165)(290,170)
\ArrowLine(315,180)(290,170)
\Vertex(315,180){1}

\ArrowLine(315,162)(315,180)
\Line(315,162)(315,150)

\Vertex(315,150){1}
\ArrowLine(275,165)(315,150)

\Vertex(290,170){1}
\Gluon(290,170)(315,162){2}{5}
\Vertex(315,162){1}

\Photon(265,165)(275,165){2}{3}
\Photon(315,180)(325,180){2}{3}
\Photon(315,150)(325,150){2}{3}

\Vertex(65,105){1}
\Vertex(105,120){1}
\ArrowLine(105,90)(105,120)
\Vertex(105,90){1}
\Line(65,105)(85,113)
\ArrowLine(105,120)(85,113)

\ArrowLine(85,97)(105,90)
\Line(85,97)(65,105)

\Vertex(85,97){1}
\Gluon(85,113)(85,97){2}{3}
\Vertex(85,113){1}

\Photon(55,105)(65,105){2}{3}
\Photon(105,120)(115,120){2}{3}
\Photon(105,90)(115,90){2}{3}

\Vertex(135,105){1}
\Line(135,105)(155,113)
\ArrowLine(155,113)(175,120)

\ArrowLine(175,90)(155,97)
\Line(155,97)(135,105)

\Vertex(155,97){1}
\Gluon(155,113)(155,97){2}{3}
\Vertex(155,113){1}

\Vertex(175,120){1}

\ArrowLine(175,120)(175,90)
\Vertex(175,90){1}
\Photon(125,105)(135,105){2}{3}
\Photon(175,120)(185,120){2}{3}
\Photon(175,90)(185,90){2}{3}

\Vertex(205,105){1}
\ArrowLine(245,120)(205,105)
\Vertex(245,120){1}
\ArrowLine(245,90)(245,120)

\GlueArc(245,105)(7,90,270){2}{4}
\Vertex(245,97){1}
\Vertex(245,113){1}

\Vertex(245,90){1}
\ArrowLine(205,105)(245,90)
\Photon(195,105)(205,105){2}{3}
\Photon(245,120)(255,120){2}{3}
\Photon(245,90)(255,90){2}{3}

\Vertex(275,105){1}
\ArrowLine(275,105)(315,120)
\Vertex(315,120){1}
\ArrowLine(315,120)(315,90)
\Vertex(315,90){1}
\ArrowLine(315,90)(275,105)

\GlueArc(315,105)(7,90,270){2}{4}

\Photon(265,105)(275,105){2}{3}
\Photon(315,120)(325,120){2}{3}
\Photon(315,90)(325,90){2}{3}

\Vertex(315,97){1}
\Vertex(315,113){1}


\Vertex(65,45){1}
\ArrowLine(65,45)(105,60)
\Vertex(105,60){1}
\ArrowLine(105,60)(105,30)
\Vertex(105,30){1}
\ArrowLine(105,30)(65,45)
\Photon(55,45)(65,45){2}{3}
\Photon(105,60)(115,60){2}{3}
\Photon(105,30)(115,30){2}{3}

\GlueArc(85,37)(7,-12,159){2}{4}
\Vertex(80,39){1}
\Vertex(91,35){1}

\Vertex(135,45){1}
\ArrowLine(175,60)(135,45)
\Vertex(175,60){1}
\ArrowLine(175,30)(175,60)
\Vertex(175,30){1}
\ArrowLine(135,45)(175,30)
\Photon(125,45)(135,45){2}{3}
\Photon(175,60)(185,60){2}{3}
\Photon(175,30)(185,30){2}{3}
\GlueArc(155,37)(7,-14,157){2}{4}
\Vertex(150,39){1}
\Vertex(161,35){1}

\Vertex(205,45){1}
\ArrowLine(245,60)(205,45)
\Vertex(245,60){1}
\ArrowLine(245,30)(245,60)
\Vertex(245,30){1}
\ArrowLine(205,45)(245,30)
\Photon(195,45)(205,45){2}{3}
\Photon(245,60)(255,60){2}{3}
\Photon(245,30)(255,30){2}{3}
\GlueArc(225,52)(7,-163,22){2}{4}
\Vertex(218,50){1}
\Vertex(232,55){1}

\Vertex(275,45){1}
\ArrowLine(275,45)(315,60)
\Vertex(315,60){1}
\ArrowLine(315,60)(315,30)
\Vertex(315,30){1}
\ArrowLine(315,30)(275,45)
\Photon(265,45)(275,45){2}{3}
\Photon(315,60)(325,60){2}{3}
\Photon(315,30)(325,30){2}{3}
\GlueArc(295,52)(7,-163,22){2}{4}
\Vertex(288,50){1}
\Vertex(302,55){1}

\end{picture}
\end{center}

\vspace{1cm}
\begin{center}
{ Two-loop QCD diagrams contributing 
 to  $\langle VVA \rangle$ correlator }
\end{center}


Our procedure of treating $\gamma_5$ is similar to the one used 
in~\cite{Jones:1982zf}. We write down all
fermion loops starting with the axial-vector vertex, and then perform
Feynman integrals and Dirac algebra without assuming any property of
$\gamma_5$ at all. In this way all diagrams will be expressed in terms
of traces of 10 combinations of $\gamma$ matrices: 

\begin{eqnarray}
\begin{array}{lll}
4i A_1 = \gamma_{\rho} \gamma_5 \gamma_{\mu} \gamma_{\nu} \hat{q}_2 ,
& 4i A_2 = \gamma_{\rho} \gamma_5 \hat{q}_1 \gamma_{\mu} \gamma_{\nu} , 
& 4i A_3 =  q_2^{\mu} \gamma_{\rho} \gamma_5  \hat{q}_1 \gamma_{\nu}
\hat{q}_2, \\
 4i A_4 =  q_1^{\mu} \gamma_{\rho} \gamma_5  \hat{q}_1 \gamma_{\nu} \hat{q}_2,  
&4i A_5 = - q_2^{\nu} \gamma_{\rho} \gamma_5 \hat{q}_1 \gamma_{\mu} \hat{q}_2,  
&4i A_6 = - q_1^{\nu} \gamma_{\rho} \gamma_5 \hat{q}_1 \gamma_{\mu}\hat{q}_2,
 \\
4 i A_7 = \gamma_{\rho} \gamma_5 \hat{q}_1,~~~
&4 i A_8 = \gamma_{\rho} \gamma_5 \hat{q}_2,~~~~
&4 i A_9 = \gamma_{\rho} \gamma_5 \gamma_{\mu},\\
4iA_{10}= \gamma_{\rho} \gamma_5 \gamma_{\nu} & & 
\end{array}
\end{eqnarray}


with $\hat{q} \equiv q_\mu \gamma^\mu$. The prescription is sufficient
to enable us to arrive at expressions in
front of $A_1,{\ldots} A_{10}$ which have finite limits as
$d\rightarrow 4 $.  After this the usual formulae
\begin{eqnarray}
&&
{\rm Tr} [\gamma_5 \gamma_{\alpha}\gamma_{\beta}\gamma_{\mu}\gamma_{\nu}]
=4i\veps_{\alpha \beta \mu \nu}\;,\;\;
{\rm Tr} [\gamma_5 \gamma_{\alpha}\gamma_{\beta}] = 0
\end{eqnarray}
valid in $d=4$ dimensions were used. In our convention
$\veps_{0123}=+1$ and $(1-\gamma_5)/2$ projects to left--handed
fermion fields.

Tensor integrals were expressed in terms of integrals with different
shifts of the space-time dimension~\cite{Tarasov:1996br}.  All scalar
integrals were reduced to 6 master integrals by using the Gr\"obner
basis technique proposed in~\cite{Tarasov:1998nx}.  The expressions
for the individual diagrams are sums over 15 terms which are
combinations of the 6 basis integrals

\begin{eqnarray}
I_2^{(d)}(q_j^2) &=& \int  \frac{d^dk_1}{[i
\pi^{d/2}]}\frac{1}{k_1^2(k_1-q_j)^2},
\nonumber \\
I_3^{(d)}(q_1^2,q_2^2,q_3^2) &=& \int  \frac{d^dk_1}{[i
\pi^{d/2}]}\frac{1}{k_1^2(k_1-q_1)^2(k_1-q_2)^2}
\nonumber \\
J_3^{(d)}(q_j^2) &=& \int \int \frac{d^dk_1 d^dk_2}{[i
\pi^{d/2}]^2}\frac{1}{k_1^2(k_1-k_2)^2(k_2-q_j)^2},
\nonumber \\
R_1(q_1^2,q_2^2,q_3^2)&=&\int \int \frac{d^dk_1 d^dk_2}{[i
\pi^{d/2}]^2}\frac{1}{k_1^2(k_1-k_2)^2(k_2-q_1)^2(k_2+q_2)^2}
\nonumber \\
R_2(q_1^2,q_2^2,q_3^2) &=& \int \int \frac{d^dk_1 d^dk_2}{[i\pi^{d/2}]^2}
\frac{1}{k_1^4(k_1-k_2)^2(k_2-q_1)^2(k_2+q_2)^2},
\nonumber \\
P_5(q_1^2,q_2^2,q_3^2)&=&\int \int \frac{d^dk_1 d^dk_2}
{[i\pi^{d/2}]^2}\frac{1}{k_1^2k_2^2(k_1-k_2)^2(k_1-q_1)^2(k_2+q_2)^2}.
\label{planarbasis}
\end{eqnarray}

and multiplied by ratios of polynomials in momenta and $d$: 
\begin{equation}
D_j = \sum_{k=1}^{15} M_k \frac{P_k(q_r^2,d)}{Q_k(q_s^2,d)}\epo
\end{equation}

The momentum dependence of the denominators turns out to be rather simple:
\begin{equation}
Q_k(q_s^2,d) = Q(d)(q_1^2)^{a_k}(q_2^2)^{b_k}(q_3^2)^{c_k}\Delta^{e_k}
\end{equation}
where $a_k,b_k,c_k,e_k$,  are some numbers,
$Q(d)$ is a polynomial in $d$ and
\begin{equation}
\Delta= q_1^4 + q_2^4 + q_3^4 - 2q_1^2q_2^2 - 2q_1^2q_3^2 - 2q_2^2q_3^2\epo
\end{equation}

The integrals (\ref{planarbasis}) form a complete set of master
integrals needed for the calculation of massless vertex diagrams with
planar topology.

The  sum of all diagrams turns out to be  gauge parameter
independent.

In the Feynman gauge at $q_3=0$ and for arbitrary $d$, the results of
our calculation are in agreement with the ones presented
in~\cite{Jones:1982zf} diagram by diagram.

By applying the prescription outlined above for the evaluation 
of individual diagrams, the tensor structures
\begin{eqnarray}
N_1= q_{1}^{\rho}  \veps_{\alpha \beta \mu \nu}\; ,
N_2= q_{2}^{\rho}  \veps_{\alpha \beta \mu \nu}\; ,
\end{eqnarray}
which appear in the Ansatz (\ref{calw}) are actually absent. In place
of $N_1,N_2$, two other tensor structures are present which we eliminate
by means of the Schouten identities:
\begin{eqnarray}
&&q_{1}^{\alpha}q_{2}^{\beta} q_{2}^{ \nu} \veps_{\alpha \beta \mu \rho} 
  =  +q_{1}^{\alpha}q_{2}^{\beta} q_{2}^{\rho}  \veps_{\alpha \beta \mu \nu} 
     +  q_{1}^{\alpha}q_{2}^{\beta} q_{2}^{ \mu} \veps_{\alpha \beta \nu \rho} 
     - q_2^2 q_{1}^{\alpha} \veps_{\alpha \mu \nu \rho}
     + q_{2}^{\alpha} q_1q_2 \veps_{\alpha \mu \nu \rho},
\nn \\
&& q_{1}^{\alpha}q_{2}^{\beta}    q_{1}^{ \mu} \veps_{\alpha \beta \nu \rho}  = 
- q_{1}^{\alpha}q_{2}^{\beta}   q_{1\rho} \veps_{\alpha \beta \mu\nu} 
 +q_{1}^{\alpha}q_{2}^{\beta}  q_{1}^{\nu} \veps_{\alpha \beta \mu \rho}
 +q_{1}^{\alpha} q_1q_2 \veps_{\alpha \mu \nu \rho}
 -q_{2}^{\alpha} q_1^2 \veps_{\alpha \mu \nu \rho}\;.
\end{eqnarray} 
Reshuffling terms in this way allows us to express each diagram in 
terms of the tensor structures introduced in (\ref{calw}) exhibiting
manifestly the vector current conservation.


\section{Results and Discussion}

Including one-- and two--loop contributions, we may represent the 
form-factors in the form
\bea
w_T^{(\pm)}(q_1^2,q_2^2,q_3^2) &=&
 n_f\:N_c\: w_{1,T}^{(\pm)}(q_1^2,q_2^2,q_3^2)~+~ a~ n_f\:N_c\: C_2(R)\:  w_{2,T}^{(\pm)}(q_1^2,q_2^2,q_3^2)
\crn 
w_L(q_1^2,q_2^2,q_3^2) &=&   
 n_f\:N_c\: w_{1,L}(q_1^2,q_2^2,q_3^2)~+~ a~ n_f\:N_c\: C_2(R)\:  w_{2,L}(q_1^2,q_2^2,q_3^2)
\label{formfactor}
\eea
where 
\begin{equation}
a = \frac{\alpha_s}{4\pi}=\frac{g^2}{16\pi^2}.
\end{equation}
includes the QCD coupling $\alpha_s$, $g$ as usual is the gauge coupling,
$n_f$ is the number of flavors and $N_c$ the number of colors.
The quarks are in the fundamental representation $R$ and the
corresponding group theory factor is
given by
\be
C_2(R) I = R^{a}R^{a}~~,~~~C_2(R)=4/3 ~~\mathrm{for \ QCD} \epo
\ee

We have been working in the $\overline{\rm MS}$ renormalization
scheme. The singlet axial current $J_{\rho}^5=A_\rho$ is non-trivially
renormalized because of the axial anomaly. It is
known~\cite{Trueman:1979en} that in addition to the standard
ultraviolet renormalization constant $Z_{\overline{\rm MS}}$ which
reads $Z_{\overline{\rm MS}}=1$ in our case (as $Z_{\overline{\rm
MS}}-1=O(a^2)$), one has to apply a finite renormalization constant
$Z_5$ such that renormalized and bare currents are related as:
\begin{equation}
(J_{\rho}^5 )_r = Z_5 Z_{\overline{\rm MS}}\: (J_{\rho}^5 )_0.
\end{equation}
The counterterms coming from the wave function renormalization of
quarks and ultraviolet renormalization of the axial and vector
currents cancel.  The finite renormalization constant is known at the
three-loop level~\cite{Larin:1993tq}.  For our calculations we take
\begin{equation}
Z_5=1 - 4C_2(R)\:a \epo
\end{equation}

The result of the two--loop calculation after adding all diagrams is surprisingly simple and,
normalized according to (\ref{formfactor}), is given by 
\begin{eqnarray}
q_3^2 w_{2,L}(q_1^2,q_2^2,q_3^2) &=& -8
\\
{\widetilde{w}}_{2,T}^{(-)}(q_1^2,q_2^2,q_3^2) &=& -  w_{2,T}^{(-)}(q_1^2,q_2^2,q_3^2),
\\
q_3^2 \Delta^2  w^{(-)}_{2,T}(q_1^2,q_2^2,q_3^2)
&=& 8 (x-y)\Delta
+8(x-y)(6xy + \Delta)\Phi^{(1)}(x,y)
\nonumber
\\
&-&4 [18 x y + 6 x^2-6 x + (1+x+y)\Delta)] L_x
\nonumber
\\
&+& 4[18 x y + 6 y^2-6 y + (1+x+y)\Delta)] L_y
\\
\nonumber 
\\
q_3^2 \Delta^2  w^{(+)}_{2,T}(q_1^2,q_2^2,q_3^2)&=&8[6xy + (x+y)\Delta]\Phi^{(1)}(x,y)
+8\Delta
\nonumber 
\\
&-&4 [6 x+ \Delta ](x-y-1) L_x
\nonumber \\
&+&4 [6 y+ \Delta] (x-y+1) L_y
\end{eqnarray}

with

\begin{equation}
L_x = \ln x,~~~~
L_y = \ln y,~~x=\frac{q_1^2}{q_3^2}~~y=\frac{q_2^2}{q_3^2}\epo
\end{equation}

The explicit expression for $\Phi^{(1)}$ may be found 
in~\cite{Usyukina:1992jd}:
\be
\label{Phi1}
\Phi^{(1)} (x,y) = \frac{1}{\lambda} \left\{ \frac{}{}
2 \left( \Li{2}{-\rho x} + \Li{2}{-\rho y} \right)
+ \ln\frac{y}{x} \ln{\frac{1+\rho y}{1+\rho x}}
+ \ln(\rho x) \ln(\rho y) + \frac{\pi^2}{3}
\right\} ,
\ee
where
\be
\label{lambda}
\lambda(x,y) \equiv \sqrt{\Delta} \; \; \; ,
\; \; \; \rho(x,y) \equiv 2 \; (1-x-y+\lambda)^{-1}, ~~\Delta=(1-x-y)^2 - 4 x y \;.
\ee

The comparison with the results of the one-loop calculation reveals that

\be
{{\cal W}}_{\mu\nu\rho}(q_1,q_2)\left|_{two-loop}\right.=
4C_2(R) a {{\cal W}}_{\mu\nu\rho}(q_1,q_2)\left|_{one-loop}\right.
\ee

Multiplying the sum of one- and two-loop terms by the finite factor 
$Z_5$ we arrive at
\be
{{\cal W}}_{\mu\nu\rho}(q_1,q_2)
 = {{\cal W}}_{\mu\nu\rho}(q_1,q_2)\left|_{one-loop}\right.
\ee
This is the non-renormalization theorem for the full off shell
correlator at two--loops. While our calculation confirms the relations
(\ref{theorems}) derived in~\cite{KPPdR04} and the non-renormalization
theorem (\ref{nonrentrans}) found in~\cite{Vainshtein03,CMV03}, these findings are not
sufficient to explain our result valid for generic momenta.

Taking into account the rather non-trivial momentum dependence of the
form-factors it is very tempting to suggest that it could hold to all
orders of perturbation theory because of the topological nature of the
anomaly, for example. 

We would like to stress that the surprising relation could be
discovered only by keeping the general non-trivial momentum
dependence.  The anomalous three point correlator exhibits an
unusually simple structure, while contributions from individual
diagrams are very unwieldy.  One can expect similar effects for other
anomalous correlators. Since at the order considered the QCD
calculation is essentially a QED calculation, it is highly non-trivial
whether this carries over to higher orders. For the $\VVdA$ anomalous
correlator a large number of two--loop calculations have been
performed (\cite{Adler:1972zd}-\cite{Bos:1992nd}), mainly in QED
and we refer to the comprehensive review by Adler~\cite{Adler:2004ih}
and the references therein.\\

In electroweak SM calculations one would a priori expect that
remormalizing parameters and fields would be sufficient for
renormalizing the SM. Our calculation shows that on top of the
standard renormalization, it is mandatory to renormalize the anomalous
currents $J^5_\rho$ by the finite renormalization factor $Z_5$ because
the lepton currents and the quark currents pick different
$Z$--factors and if they are not renormalized away the anomaly
cancellation and hence renormalizability obviously would get
spoiled. As pointed out by Adler and many others~\cite{Adler:2004ih}
the point is there exists a renormalization scheme for which the
one--loop anomaly is exact. Only in this scheme anomaly cancellation
and thus renormalizability will carry over to higher orders in the
SM. Our result shows that due to the necessity of renormalizing away
possible higher order contributions from the anomaly also the
non-anomalous transversal contributions are affected. We have shown
that at least at two--loops the entire contribution gets renormalized
away in the zero mass limit.\\[4mm]

{\bf Acknowledgments}\\

This work was supported by DFG Sonderforschungsbereich Transregio 9-03
and in part by the European Community's Human Potential Program under
contract HPRN-CT-2002-00311 EURIDICE and the TARI Program under contract 
RII3-CT-2004-506078. We are grateful to Harvey Meyer for carefully reading
the manuscript.

\end{document}